\documentclass[
    final            % use final for the camera ready runs
  ,numberedheadings % uncomment this option for numbered sections
  ]
  {aipproc}

\layoutstyle{6x9}

\begin{document}

%\title{Adaptive Time-evolution of one-dimensional Quantum Many Body
%  Systems using the Density Matrix Renormalization Group -- a Krylov
%  Space Approach} 
\title{Time evolution of one-dimensional Quantum Many Body Systems} 

\author{Salvatore R. Manmana}{
  address={Institut f\"ur Theoretische Physik III, Universit\"at
  Stuttgart, Pfaffenwaldring 57/V,\\ D-70550 Stuttgart, Germany}
 ,altaddress={AG Vielteilchennumerik, Fachbereich Physik,
  Philipps-Universit\"at Marburg, \\D-35032 Marburg, Germany}
}

\author{Alejandro Muramatsu}{
address={Institut f\"ur Theoretische Physik III, Universit\"at
  Stuttgart, Pfaffenwaldring 57/V,\\ D-70550 Stuttgart, Germany}
}

\author{Reinhard M. Noack}{
  address={AG Vielteilchennumerik, Fachbereich Physik,
  Philipps-Universit\"at Marburg, \\D-35032 Marburg, Germany}
  %,altaddress={<author1 address>} % additional visiting address
}

\begin{abstract}
%RMN Only little is known about 
The level of current understanding of the physics of time-dependent
strongly correlated quantum systems is far from complete, principally 
due to the lack of effective controlled approaches. 
Recently, there has been progress 
%RMN in this direction
in the development of approaches for one-dimensional systems. 
We describe recent developments in the
construction of numerical schemes for general (one-dimensional)
Hamiltonians: in particular, schemes based on exact 
diagonalization techniques and on the density matrix renormalization
group method (DMRG).
We present preliminary results for spinless fermions with
nearest-neighbor-interaction and investigate their accuracy by
comparing with exact results. 
\end{abstract}

\maketitle

%%%%%%%%%%%%%%%%%%%%%%%%%%%%%%%%%%%%%%%%%%%%
%% MAINMATTER
%%%%%%%%%%%%%%%%%%%%%%%%%%%%%%%%%%%%%%%%%%%%

\section{Introduction}
\label{sec::intro}
The time evolution of strongly interacting quantum many body systems 
is one of the most challenging experimental and theoretical problems
in physics.
Recently, there has been progress in the investigation of the 
non-equilibrium time evolution of quantum systems in optical
lattices \cite{Bloch} which has spurred theoretical interest in the
time evolution of many body systems.
The analytical treatment of such systems in equilibrium is
restricted to only a few models, but in low dimensions efficient
{\it numerically exact} methods have been developed and successfully
applied to a variety of models in recent years. 

The perturbative Keldysh formalism can be
applied to investigate the non-equilibrium
behavior of such systems analytically \cite{keldysh_review}. 
Also, nonperturbative
approximation schemes basing on functional integral techniques are
under development in the context of high-energy physics \cite{dynamics_QFT}.
Due to their perturbative or approximate character, these approaches
cannot treat systems for arbitrary parameter values.
%arbitrary physical situations. %srm 
%cannot treat systems at large interaction strength (?). 
Therefore, development of  numerical tools which 
%RMN can handle arbitrary interaction strength 
do not have such limitations
is crucial. 
In this contribution, we will focus on one-dimensional systems because
the %RMN effective and controlled 
techniques developed 
in the past decade work very well for them, whereas it is rather
difficult to apply these methods with the same accuracy and efficiency
to higher dimensional systems.

A number of different numerical methods for the
investigation of one-dimensional strongly correlated quantum systems
exist.  
The most important methods are Quantum Monte-Carlo (QMC)
\cite{QMC_review1,QMC_review2}, exact diagonalization
(ED) \cite{ED_review}, and the density-matrix renormalization group
(DMRG) \cite{dmrg_prl, dmrg_prb, dmrgbook, uli_rmp}.
Some approaches to time evolution using QMC are
known \cite{qmc_mak_timeevolve,qmc_egger_timeevolve} and others are under
development \cite{QMC_timeevolve}, but the
numerical calculations are difficult to control. 
Therefore, we focus on the exact diagonalization and on the DMRG.   
Although it is possible to calculate all desired
quantities using the ED for small system sizes, it is necessary to use
the more involved DMRG method in order to reach 
%RMN system sizes relevant to experiment.
sufficiently large system sizes to carry out a controlled finite-size
scaling to the thermodynamic limit 
or to describe experiments with a large but finite number of sites,
which are, e.g., realizable in optical lattices. %srm
In this contribution, we present results for the time evolution of a
one-dimensional system obtained using an ED approach and use them for
testing different DMRG approaches to the time evolution of the same
system.  

The main difficulty in calculating the time evolution using the DMRG
is that the effective basis determined at the beginning of the
time evolution is not able, in general, to represent the state well at
later times \cite{luo:049701} because it %RMN the approximate effective basis
covers a subspace of the system's total Hilbert space which is not
appropriate to properly represent the state at the next time step.
As will be shown in Sec. \ref{sec::adaptive_tDMRG}, it is possible that the
representation of the time-dependent wave function very soon becomes
quite bad.
%srm: here we should shorten a bit:
%The easiest way to overcome this difficulty is to store {\it all} the
%$N+1$ states $|\psi(t_i)\rangle$ obtained during the time evolution at
%the time steps $t_0, t_1, \ldots, t_N$ and mix them into the density
%matrix, i.e., to perform the DMRG procedure with these $N+1$ target
%states \cite{luo:049701,schmitteckert:121302}. 
%This makes the DMRG inefficient and the calculation very expensive.
%RMN An elegant possibility to really adapt the density-matrix basis and
%avoid this big amount of target states is using a Trotter-Suzuki
%approximation of the time-evolution operator in the way developed in
%References  \cite{vidal:040502,white:076401, Uli_timeevolve}. 
%Recently, an alternative method based on a Trotter-Suzuki
%approximation of the time-evolution operator has been developed
%\cite{vidal:040502,white:076401, Uli_timeevolve}.
%This method adapts the density-matrix basis to the new time step
%incrementally as the 
%local %srm
%pieces of the Trotter-Suzuki-decomposed
%time-evolution operator are applied, thus avoiding targeting a large
%number of states.
% 
It is necessary either to mix {\it all} time steps $|\psi(t_i)\rangle$
  into the density-matrix \cite{luo:049701,schmitteckert:121302}, or
  to {\it adapt} the density matrix.
An approach for adaptive time-evolution basing on the Trotter-Suzuki
decomposition of the time-evolution operator was developed in
Refs. \cite{vidal:040502,white:076401, Uli_timeevolve}.  
However, the method is restricted to systems with nearest-neighbor
terms in the Hamiltonian only. 
In our approach, we use a general scheme %RMN intimately connected 
closely related to exact 
diagonalization techniques to obtain the next time step during the
time evolution, combined with an adaption scheme
recently proposed by White \cite{steve_leiden,white_timeevolve_newpaper}.
%presented by
%Steve White in ``the DMRG workshop'' in Leiden this
%year \cite{steve_leiden}. 
Using this more general approach, it is possible to treat Hamiltonians
containing arbitrary terms as long as the system can be efficiently handled
with standard DMRG procedures.

The article is organized as follows. 
First, we briefly review the DMRG method in 
Sec.\ \ref{sec::dmrg_method}. 
In Sec.\ \ref{sec::approaches}, we discuss general schemes for
obtaining the time evolution of arbitrary quantum systems.   
In Sec.\ \ref{sec::ed}, we focus on strongly correlated systems
and present results calculated using an exact diagonalization scheme.
Subsequently, in  Sec.\ \ref{sec::adaptive_tDMRG}, we discuss
approaches and preliminary results obtained with non-adaptive and
adaptive time evolution using the DMRG method and compare the results
with the outcomes of the exact diagonalization approach.
Finally, we discuss the results
obtained and future work in Sec.\ \ref{sec::discussion}.

\begin{figure}
\includegraphics[width=0.95\textwidth]{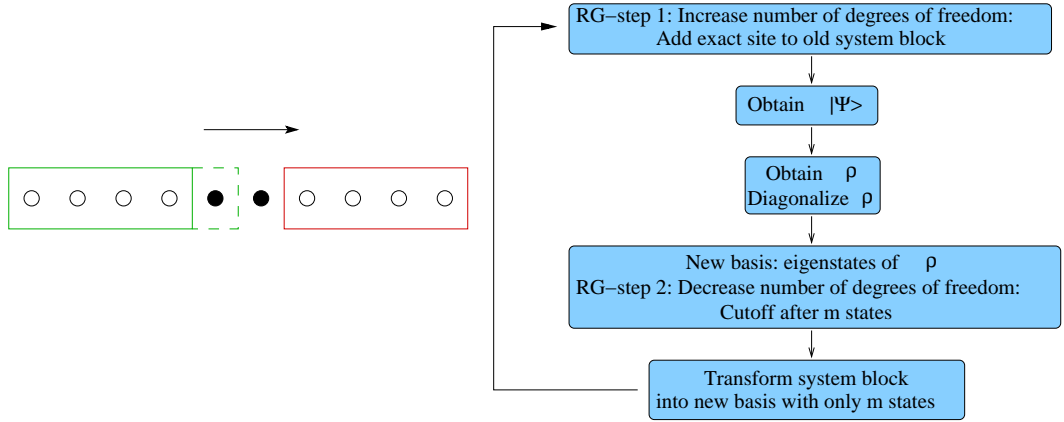}
\caption{Sketch of the lattice and the flowchart of the DMRG iteration
  scheme. The left part of the lattice is the subsystem to which a
  site is added; the ``sweep'' proceeds from left to right. The
  flowchart shows the relevant steps of the DMRG procedure. For
  further details of the method see \cite{proceedings_Vietri_DMRG}.} 
\label{fig::dmrg_flowchart} 
\end{figure}

\section{The DMRG method}
\label{sec::dmrg_method}
The DMRG method \cite{dmrg_prl,dmrg_prb} is described in detail in
another contribution in 
this volume \cite{proceedings_Vietri_DMRG}. 
However, in order to better understand the main difficulties related
to the calculation of the time evolution of an initial state
$|\psi_0\rangle$ using DMRG, we will give a short description of the
main steps of the method here. 
The basic idea of the density-matrix renormalization group method is
to represent one or more pure states of a finite system approximately
by dividing the system in two and retaining only the $m$ most highly
weighted eigenstates of the reduced density 
matrix of the partial system.
% as the basis in which the Hamiltonian
%of this subsystem is represented. 
In combination with the numerical renormalization group approach (NRG)
developed by Wilson \cite{wilson_rg} and the superblock algorithms
developed by White and Noack \cite{white_and_noack}, this leads 
to a very powerful and efficient tool for the investigation of
one-dimensional strongly correlated quantum systems on a lattice.   

%The DMRG is an iterative approach;
%srm: shorten here:
%Iterative DMRG schemes exist either to increase the system size
%(``infinite-system algorithm'') or to iteratively converge to
%a particular state (typically the ground state) on a finite system
%(``finite-system algorithm'').  
As depicted in Fig.\ 1, the key steps are to {\it increase} the
number of degrees of freedom of the partial system by adding sites,
then to {\it decrease} the number of degrees freedom by retaining
states below a cutoff.
In this way, the method carries out a renormalization group procedure
closely related to Wilson's NRG.  

%The system is divided into two subsystems and
%the number of degrees of freedom is increased by
%adding a site to one of the two subsystems,
In the first step of the algorithm, in which a site is added to one of
the subsystems, 
the site-Hamiltonian is represented {\it exactly} in the site's
many-body basis.
This basic feature is exploited by the Trotter-Suzuki approach to
time evolution \cite{white:076401, Uli_timeevolve} to efficiently
apply the 
local %srm
time-evolution operator in this part of the lattice. 
%srm: shorten here:
%While the Hamiltonian of the subsystem to which the site is added may
%in principle be represented in the exact many-body basis, it usually (i.e. after the 
%first iteration step) 
The subsystem's Hamiltonian is usually represented in an {\it
  efficient} reduced basis built up from the $m$ most important
  eigenstates of the subsystem's reduced density-matrix.
%Note that due to the truncation the basis is incomplete. 
In the second step, the states one is interested in are
obtained. 
These states are called ``target states''.
In the original ground-state algorithm, these are the ground
state and the few lowest lying 
excited states of the system, which are obtained by diagonalizing the
Hamiltonian of  
the {\it total} system, e.g. by using the L\'anczos diagonalization
algorithm described in Sec.\ \ref{sec::ed}. 
However, one is not restricted to eigenstates of the Hamiltonian, and in
this step may obtain alternative states. 
%srm This will become crucial for the time-evolution algorithms
%presented below. 
These properties are crucial for the time-evolution algorithms. 
The main problem is that the time-evolved state is not given by
solving an eigenvalue problem. 

%srm In the third step, the reduced density matrix of the extended
%subsystem is obtained in the usual way, i.e.,
In the third step, the new effective basis is obtained by diagonalizig
the reduced density matrix of the extended subsystem given by
\begin{equation}
\rho_{\rm subsystem} = {\rm Tr}_{\rm rest} \left(\sum\limits_i n_i | \psi_i
\rangle \langle \psi_i |  \right) \ , \quad \sum n_i = 1 \; ,
\end{equation}
where the sum goes over all target states.
%srm In the fourth step, $\rho_{\rm subsystem}$ is diagonalized and
%the new effective basis formed by keeping only the $m$ eigenstates
%with the highest eigenvalues.
In step four, only the $m$ eigenstates with the largest eigenvalues
are kept. 
The operators needed to represent the subsystem's Hamiltonian, to form
the pieces of the Hamiltonian connecting subsystems, and 
%srm to carry out measurements 
to calculate observables 
are transformed into this new reduced basis. 
This effective Hamiltonian of the subsystem is now the starting point
for step one of the next iteration. 
%RMN For further methodical and implementation details, see the other
%contribution in this edition.
In this way, every iteration step improves the accuracy of the
obtained eigenstates and energies
%RMN This is mainly due to the increasing ability of the effective basis to
%accurately represent the target states in the reduced Hilbert space 
%achieved in the course of the iteration procedure.
%SRM by improving the reduced Hilbert space as a representation of
%the target states.
by improving the reduced basis used for the representation of the
target states. 

%This is reinforced by the properties of the eigensolvers like the
%Lanczos diagonalization, which can take the ground state  from
%the previous iteration as input and obtain new eigenstate(s) which are 
%closer to the true solution of the full problem (similar to
%predictor--corrector methods) from it. 
%srm: shorten here:
%In order to target an extremal eigenstate such as the ground state, an
%iterative diagonalization 
%routine such as the L\'anzcos or the Davidson algorithm is carried out
%on the Hamiltonian within the reduced Hilbert space.
%The time-evolved state, however, is not the solution of an
%eigenproblem. 
%srm The main problem for the time-evolution is that the time-evolved
%state is not given by solving an eigenproblem.
Since both the Hamiltonian and the wavefunction $|\psi(t)\rangle$ at
time $t$ are represented in an incomplete basis, the result for the
next time step $|\psi(t+dt)\rangle$ will 
%RMN also be erroneous.
have additional errors because the reduced basis is not an optimum
representation for this state. 
In order to minimize these errors, it is necessary to form 
a density matrix whose $m$ most important eigenvectors are
``optimal'' for the representation of the state $|\psi(t)\rangle$, as
well as for $|\psi(t+dt)\rangle$ in the reduced Hilbert space. 
In Sec. \ref{sec::adaptive_tDMRG} we will discuss ways of %RMN approaches for
dealing with this problem. 

\section{Approaches to time evolution}
\label{sec::approaches}

\begin{figure}
\includegraphics[width=0.95\textwidth]{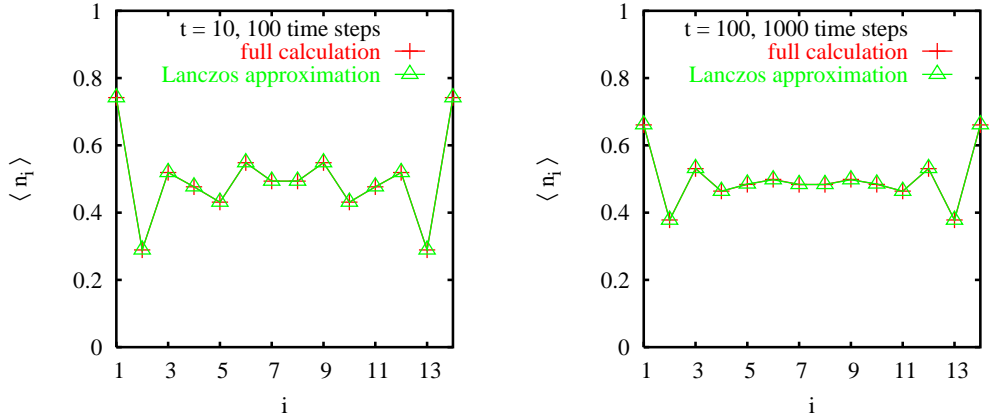}
\caption{Comparison of the time evolution of the local density
  $\langle n_i \rangle$ using full diagonalization
  and the L\'anczos-approach described in Sec.\ \ref{sec::ed} for a
  system of spinless fermions on 14 lattice sites with open boundary
  conditions and interaction $V(t=0) = 0.5, \, V(t>0)=2.5$. $m_L
  = 5$ L\'anczos vectors were kept. As can be seen, the error is
  negligible even after long times.}  
\label{fig::compare_exact_lanczos}
\end{figure}

The dynamics of a quantum system is given by the solution of the
time-dependent Schr\"odinger equation
\begin{equation}
i \hbar\frac{\partial}{\partial \, t} | \psi(t)\rangle = \hat{H} |
\psi(t) \rangle \; .
\end{equation}
This equation is a first-order ordinary differential equation (ODE)
which can be numerically solved directly by approximate 
integration schemes such as the %$4^{\rm th}$ order 
Runge-Kutta method \cite{numerical_recipes}.
%\begin{eqnarray*}
%|\psi(t+dt)\rangle &\approx& |\psi(t)\rangle + \frac{1}{6} \left(
% |k_1\rangle + 2 \left( |k_2\rangle + |k_3\rangle \right) + 
% |k_4\rangle \right) \\
% |k_1\rangle &=&  dt \cdot \left(- \frac{i}{\hbar} \hat{H}(t)
% |\psi(t)\rangle\right)\\ 
% |k_2\rangle &=& dt \cdot \left(- \frac{i}{\hbar} \hat{H}(t+ \frac{1}{2} dt) 
% \left(|\psi(t) \rangle +\frac{dt}{2} |k_1\rangle
% \right)\right)\\ 
% |k_3\rangle &=&  dt \cdot \left(- \frac{i}{\hbar} \hat{H} (t+ \frac{1}{2} dt)
% \left(|\psi(t) \rangle +\frac{dt}{2} |k_1\rangle
% +\frac{dt}{2} |k_2\rangle \right)\right)\\ 
% |k_4\rangle &=&  dt \cdot \left(- \frac{i}{\hbar} \hat{H} (t+dt)
% \left(|\psi(t) \rangle +\frac{dt}{2} |k_1\rangle
% +\frac{dt}{2} |k_2\rangle + dt |k_3 \rangle \right)\right).
%\end{eqnarray*}
This method has been used within a non-adaptive
DMRG scheme in Ref.\ \cite{cazalilla:256403} to obtain
the time-dependent tunneling currents through a quantum dot connected 
to both non-interacting and interacting leads. 
This approach is a standard approach with an error $\varepsilon
\propto (dt)^4$, but does not conserve unitarity. 
%RMN It is crucial to implement the operation
%$\hat{H}(t) |\psi(t)\rangle$ efficiently.  
The crucial numerical step in this procedure is the 
$\hat{H}(t) |\psi(t)\rangle$, which must be implemented efficiently.  
An alternative implicit integration scheme conserving unitarity is the
Crank-Nicholson procedure \cite{numerical_recipes} 
\begin{equation}
|\psi(t+dt)\rangle \approx \frac{1-i\hat{H}(t) \delta t/2}{1+i\hat{H}(t)
  \delta t/2} |\psi(t)\rangle \; .
\label{eq:Crank-Nich}
\end{equation}
This approach has been used in Ref.\ \cite{Uli_timeevolve} within 
%srm adaptive (??) DMRG
a non-adaptive DMRG approach 
to obtain the time evolution of a Bose-Hubbard
system with an instantaneous change in the interaction strength at the
beginning of the time evolution. 
In this method, the most costly numerical step is the calculation of the
inverse (the denominator in Eq.\ \ref{eq:Crank-Nich}), 
which can be carried out using, e.g., a biconjugate gradient
approach \cite{Uli_timeevolve}.  
The accuracy of this operation also determines the error of this
approach. 
Other, more involved implicit and explicit integration schemes
are known (see, e.g., Ref.\ \cite{numerical_recipes}), but none of them
has been used yet within the DMRG.   

An alternative to the direct integration of the Schr\"odinger equation
is to treat the formal solution
\begin{equation}
| \psi(t)\rangle = \hat{U} | \psi(t)\rangle 
\end{equation}
directly, where $\hat{U} = e^{-i \hat{H} t / \hbar}$ is the time evolution
operator. 
When full diagonalization of the Hamiltonian is viable, the
time evolution can be expressed in the eigenbasis by  
\[
|\psi(t) \rangle = \sum\limits_n e^{-i E_n t /\hbar} |n\rangle \langle
n | \psi_0 \rangle \; ,
\]
where $E_n$ and $|n\rangle$ are the eigenvalues and eigenfunctions of
$\hat{H}$ and $|\psi_0\rangle$ is the initial state.  
However, for strongly interacting quantum systems it is not possible
to fully diagonalize $\hat{H}$ for all but the smallest system sizes,
so that an approximation to the time-evolution operator is needed.     
In the next section we present an ED approach with which such an
approximation can be obtained within the Krylov basis to very high
accuracy.
%RMN is possible to exactly obtain the time evolution of such a system.

\section{Time evolution using exact diagonalization}
\label{sec::ed}

Efficient iterative eigensolvers exist which can calculate the ground state
and low lying eigenstates of systems with Hamiltonians which can be
represented as sufficiently sparse matrices.
The full spectrum, however, cannot efficiently be obtained with these
techniques. 
%cannot give the full spectrum of the
%system's Hamiltonian, but it is possible to calculate the ground state
%and low lying eigenstates. 
One example is the 
{\it L\'anczos-procedure} \cite{Lanczos,Lanczos_book}, which is presented 
in more detail in another contribution in this volume. 
In this procedure, the vectors of the Krylov subspace, the subspace
spanned by vectors
\[
\left\{|u_0\rangle, \hat{H} |u_0\rangle, \hat{H}^2
|u_0\rangle,...,\hat{H}^n|u_0\rangle\right\}  \; ,
\]
are orthogonalized with respect to the previous two vectors of the
set, leading to the recursion relation 
\begin{equation}
| u_{j+1} \rangle = \hat{H} | u_j \rangle - \alpha_j | u_j \rangle
  -\beta_j^2 | u_{j-1} \rangle
\end{equation}
with the coefficients
\begin{equation}
\alpha_j = \frac{\langle u_j | \hat{H} | u_j \rangle}{\langle u_j |
u_j \rangle} \, , \qquad \beta_j^2 = \frac{\langle u_j | u_j
\rangle}{\langle u_{j-1} | u_{j-1} \rangle}.
\end{equation}
The Hamiltonian is then represented by the tridiagonal
matrix 
\begin{equation}
\mathbf{T}_n=
\left(
\begin{array}{ccccc}
\alpha_0 & \beta_1    &          &            &          \\
\beta_1  & \alpha_1   & \beta_2  & \mathbf{0} &          \\
         & \beta_2    & \alpha_2 & \ddots     &          \\
         & \mathbf{0} & \ddots   & \ddots     & \beta_n  \\
         &            &          & \beta_n    & \alpha_n \\
\end{array}
 \right)
\end{equation}
which can be easily diagonalized.
For a review, see Refs.\ \cite{ED_review, Lanczos_book}.
%RMN: mixed notation: what is n?
If $n$ is equal to the dimension of the Hilbert space, the approach is
equivalent to a full diagonalization of $\hat{H}$ (albeit not a useful
one due to numerical instability). 
However, it turns out that for the calculation of the ground state it
is sufficient in most cases to carry out of the order of 
$100$ such iterations.

It is possible to use a Krylov-space approach to approximate
the time-evolution operator $\hat{U}_{dt} = e^{-idt/\hbar \, \hat{H}}$
from the time step $t$ to the time step $t+dt$.
In the following, we assume that the Hamiltonian $\hat{H}$ has no
explicit time-dependence for $t>0$. % in the course of the time evolution. 
%RMN is this right?  We need to switch on H' at some time...
%srm: H means here H'... 

The time evolution through one interval $\hat{U}_{dt} | \psi\rangle$
can be approximated 
by \cite{lubich_timeevolve,19dubiousways}   
\begin{equation}
|\psi(t+dt)\rangle = e^{-idt/\hbar \, \hat{H}} \, |\psi(t)\rangle
\approx  \mathbf{V}_n(t) \; e^{-i dt/\hbar \, \mathbf{T}_n(t)} \;
\mathbf{V}_n^{T}(t) \; |\psi(t)\rangle \label{eq::lanczos_timeevolve}
\end{equation}
where $\mathbf{V}_n$ is the matrix containing all the L\'anczos
vectors $|u_j\rangle$.
The error in this approximation is given by \cite{lubich_errors}
\begin{eqnarray}
\varepsilon_n &:=& || \; |\psi(t+dt)\rangle - |\psi(t+dt)\rangle_{\rm
  approx} || \nonumber \\
&\leq& 12 \; \exp\left\{-\frac{(\varrho \, dt)^2}{16n}\right\}
 \left(\frac{e \, \varrho \, dt}{4n}\right)^n, \\
{\rm with} \quad n &\geq& \frac{1}{2} \, \varrho \, dt
 \; . \label{eq::condition_errors} 
\end{eqnarray}
Here $|| \cdot ||$ represents the euclidean norm and $\varrho =
\left| E_{\rm max} - E_{\rm min} \right|$ is the width of the
spectrum of the Hamiltonian.
This means that the number of L\'anczos vectors and the size of the
time interval $dt$ needed to obtain a given accuracy depend on the total
energy of the system. 
In general, the larger the system, the more L\'anczos vectors are
needed, assuming that $dt$ and the filling are kept fixed.
However, this is not a serious limitation because the characteristic
oscillations 
take place on timescales of the order $\tau_{\rm char} \sim
\frac{\hbar}{\varrho}$. 
In order to fully resolve the dynamics it is necessary to adjust $dt$
to be of the order of $\tau_{\rm char}$ anyways.
Together with relation (\ref{eq::condition_errors}), for typical situations
where an accuracy of, e.g., ${\rm tol} < 10^{-6}$, is required,
one finds that $n \leq 20$ is sufficient.  
Therefore, the matrices in Eq.\ (\ref{eq::lanczos_timeevolve}) are
very small. 
The most time-consuming part of the calculation is then the multiplication
$\hat{H} | u_n \rangle$ needed to carry out the recursion to obtain
the L\'anczos vectors, it is important to implement this efficiently.  
%RMN This leads us to the following integration scheme:
Putting these steps together, the full procedure reads:
\begin{enumerate}
\item Estimate the number of L\'anczos vectors $m_L$  needed for the
  given time step to obtain $\varepsilon_n < {\rm tol}$. 
\item Obtain $V_{m_L}$ and $T_{m_L}$ by performing the L\'anczos
  iteration scheme with $|\psi(t)\rangle$ as starting vector. 
\item Compute $|\psi(t+dt)\rangle = \mathbf{V}_n(t) \; e^{-i
  dt/\hbar \, \mathbf{T}_n(t)} \; \mathbf{V}_n^{T}(t) \;
  |\psi(t)\rangle$. 
\item %Perform measurements.
  Calculate observables.
\item Continue starting with step 2 and replacing $|\psi(t)\rangle$ by
  $|\psi(t+dt)\rangle$ until $t_{\rm max}$ is reached.
\end{enumerate}

\begin{figure}
\includegraphics[width=0.95\textwidth]{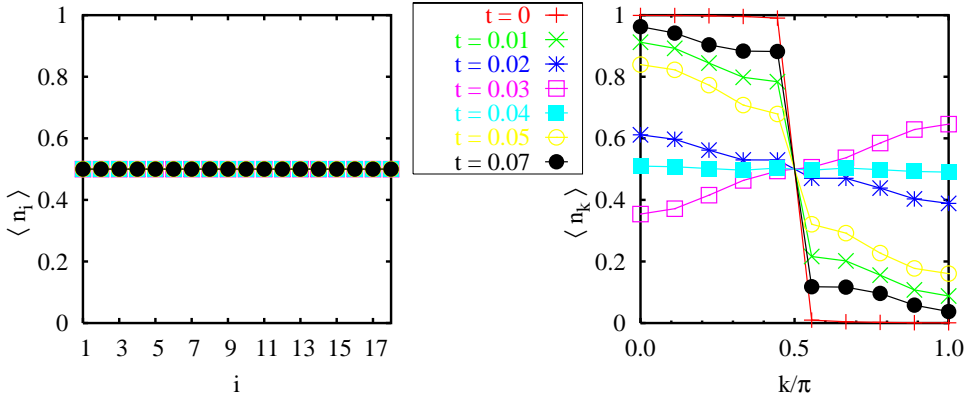}
\caption{Plot of the local density $\langle n_i \rangle$ and the
  momentum distribution $\langle n_k \rangle$ for a system of 18 sites
  with periodic boundary conditions, $V(t=0) = 0.5, \, V(t>0) = 100$
  calculated with the ED approach presented in Sec. \ref{sec::ed}. One
  can clearly observe a collapse and revival of the
  discontinuity at the Fermi edge, while $\langle n_i \rangle = 0$ for 
  all times. In this calculation, $dt =0.001$ and $m_L=10$
  L\'anczos vectors.} 
\label{fig::collapse_and_revival}
\end{figure}

Using this approach, we calculate the time evolution of a system of
spinless fermions given by the Hamiltonian 
\begin{equation}
\hat{H} = -t \sum \left(c_{j+1}^{\dagger} c_j + h.c. \right) + V \sum
n_j n_{j+1} \; ,
\end{equation}
where we change the magnitude of the interaction strength V at t=0. 
%where we adiabatically change the magnitude of the interaction
%strength $V$ at the beginning of the time evolution. 
The half-filled system is known to undergo 
a phase transition at the critical parameter value $V=2$
from a metallic ($V<2$) to a CDW insulating phase
($V>2$) \cite{spinless_fermions} in the thermodynamic limit.   
Thus, by starting with a ground state obtained for $V<2$ and applying
the time-evolution operator with an interaction strength $V>2$, one
would expect oscillations in time between these two different phases. 
In addition, one would expect time-dependent Friedel-like oscillations in the
local density $\langle n_i \rangle$ near the boundaries for finite
systems with open boundary conditions.
Such oscillations are shown in Fig.\ \ref{fig::compare_exact_lanczos}
for two different times.
In order to test the accuracy of the L\'anczos calculations, we have
compared to results obtained from full diagonalization of the Hamiltonian.
As can be seen, the error remains negligible up to the fairly long
times treated here. 

In Fig.\ \ref{fig::collapse_and_revival}, results are shown for
periodic boundary conditions for $L=18$, a system size not accessible
to full diagonalization. 
As can be seen, there is no change in the local
density, for all times. 
This is expected due to translation symmetry, and is reproduced by the
numerical calculations, which do not explicitly enforce the symmetry.

At the initial time step at temperature $T=0$, for $V<2$, the momentum
distribution $\langle n(k) \rangle$  has a singularity at the Fermi
edge in the  thermodynamic limit\cite{voit}. 
However, due to the limited system size, a
step at $k_F$ is observed rather than a singularity. 
This discontinuity vanishes and then reappears in the course of
the time evolution.
This may be interpreted as a collapse and revival of the metallic state
of the system.
Note, however, that a step at the Fermi edge in a finite one-dimensional
system does not automatically induce a metallic state in the
thermodynamic limit.
At present, it is unclear whether the singularity in the
thermodynamic limit is revived in the course of the time evolution.
This issue is under investigation and will be presented
elsewhere\cite{our_work}.  

Within full diagonalization, sizes $L \leq 14$ are
possible for this system.  
For ED, $L = 26$ can be reached on a supercomputer using a very basic
code.
By parallelizing the code and exploiting symmetries more completely,
larger system sizes could be reached. However, they would nevertheless
be small compared to the system sizes that we expect can be
treated with the time-dependent DMRG. 
%For larger system sizes comparable to experiments, DMRG is needed...

\section{Adaptive time evolution using the DMRG}
\label{sec::adaptive_tDMRG}

\begin{figure}
\includegraphics[width=0.485\textwidth]{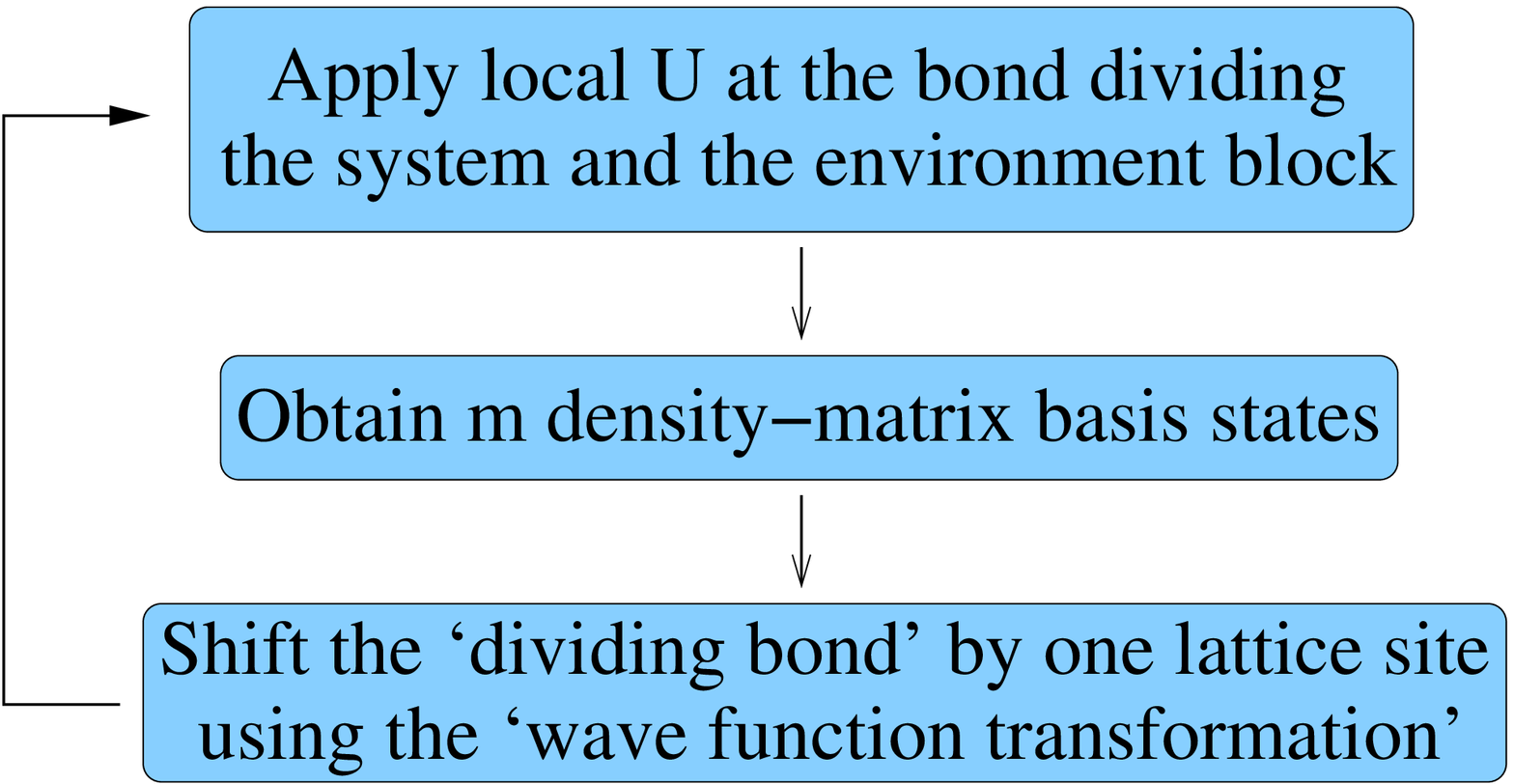}
\includegraphics[width=0.485\textwidth]{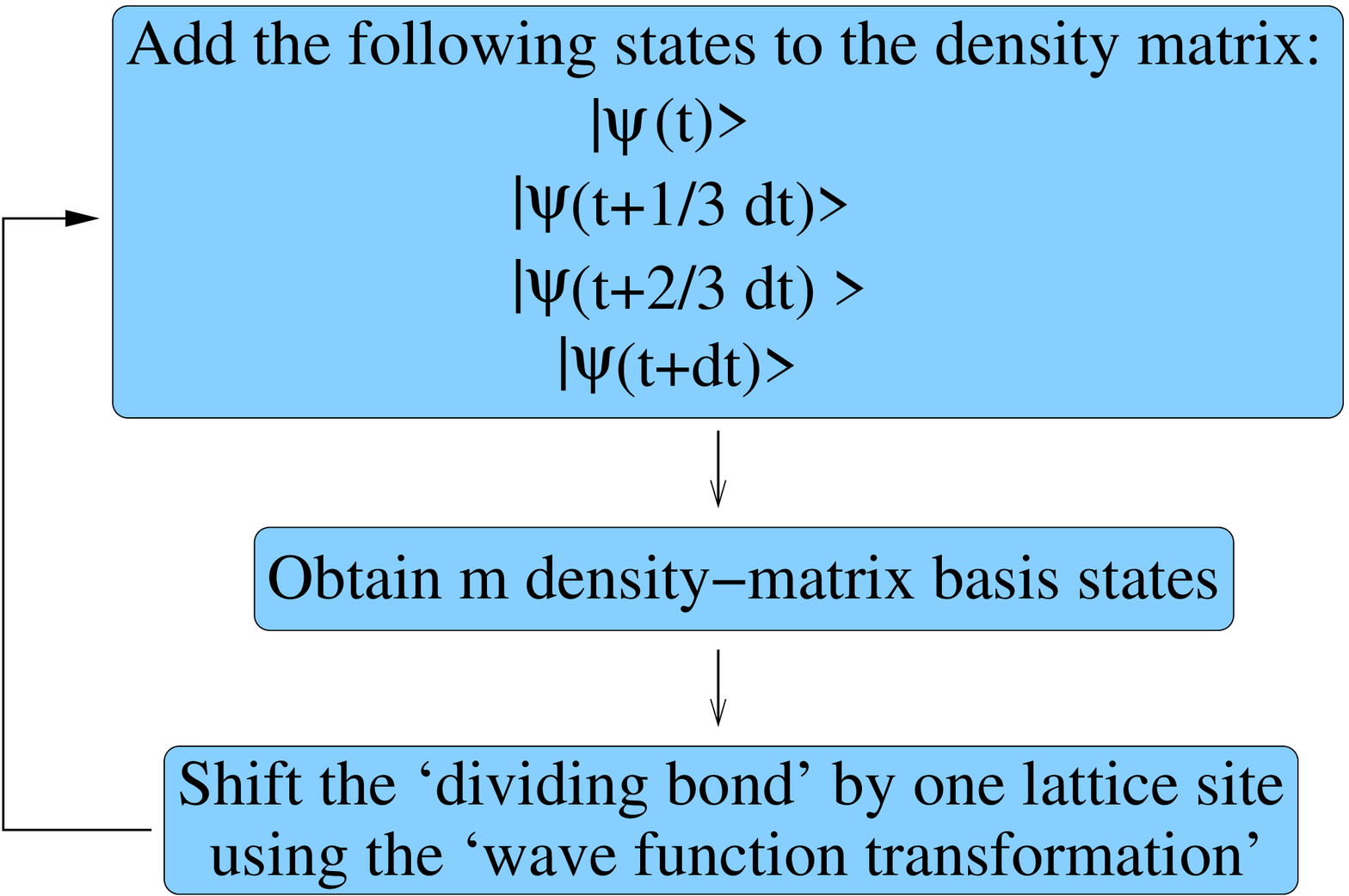}
\caption{Flowcharts of adaptive time-evolution schemes. On the left,
  the procedure used in the Trotter-approach developed in
  \cite{white:076401,Uli_timeevolve} is sketched.
  On the right, the approach presented by White in
  \cite{steve_leiden,white_timeevolve_newpaper} is shown. 
  Note that the wave-function transformation is
  needed in both approaches (see
  Refs. \cite{dmrgbook,proceedings_Vietri_DMRG}).} 
\label{fig::flowcharts_timeevolution}
\end{figure}

As pointed out in Sec.\ \ref{sec::dmrg_method}, the main problem in
doing time evolution with DMRG is the update of the density matrix
basis. 
In Fig.\ \ref{fig::flowcharts_timeevolution}, we outline the
algorithms for two different
approaches to adaptive time evolution within the DMRG.
In the first approach, a Trotter-Suzuki
decomposition \cite{trotter_suzuki} of the time-evolution operator is
used.  
The key feature of this method %RMN, described in detail in Ref.\ 
\cite{white:076401,Uli_timeevolve}
is the application of the exact local bond
time-evolution operator to the two exactly treated sites for a
particular superblock configuration, 
%in the
%lattice which are 
%represented exactly in the sites' many-body bases, 
represented in Fig.\ \ref{fig::dmrg_flowchart} by filled circles. 
Results obtained by using this approach have been presented in Refs.\
\cite{white:076401, Uli_timeevolve, gobert_timeevolve,
  corinna_timeevolve},   
an extensive  error analysis is given in \cite{gobert_timeevolve}. 
This approach is restricted to systems with 
nearest-neighbor terms in the Hamiltonian only because it relies on the
fact that the sites to which the bond operator is applied are
represented exactly.
%However, since one wants to benefit from the exact representation of
%the sites, 
In order to formulate a more general method, it is useful to
reexamine the approaches presented in Sec.\ \ref{sec::approaches}.
The most promising candidate is the L\'anczos approach, presented in
Sec.\ \ref{sec::ed}, which has  small and well-controlled errors.
%it seems convenient to try to
%formulate time evolution with DMRG using this approach.
An approach including states at all time-steps in
the density-matrix rather than basis-adaption 
has been used in Ref.\ \cite{schmitteckert:121302}. 
We now describe a L\'anczos-based approach using
{\it adaptive} time evolution within the DMRG. 
The key ingredient is the implementation of the basis adaption in
a way which represents the state $|\psi(t)\rangle$ as well as
$|\psi(t+dt)\rangle$ optimally.   
A scheme to do this has recently been proposed by White \cite{steve_leiden,white_timeevolve_newpaper}.
% on the DMRG
%workshop in Leiden this year.
In this approach, the density-matrix basis is formed by including
different time-steps {\it within} the time interval $[t,t+dt]$. 
As in standard DMRG, finite-system sweeps are performed until
convergence is achieved.
At this point, the density matrix basis has been optimized to
represent both the state at time $t$ and the state at the time
$t+dt$. 
Observables at the time step $t+dt$ can be calculated and
$|\psi(t+dt)\rangle$ is then used as the starting point for
propagation to the next time-step.
In Fig.\ \ref{fig::flowcharts_timeevolution}, we outline the
procedure. 
While we currently include time steps $t \, , t+dt/3 \, ,
t+2dt/3$, and $t+dt$ in the density matrix,
which intermediate states are optimal to include is still under
investigation. 
At present, it is unclear how the sweeping influences the
convergence behavior, although we believe that 
the accuracy of the state
$|\psi(t+dt)\rangle$ can be improved by additional sweeping. 
Also, the errors in the reduced basis will accumulate with time. 
A more thorough error analysis and an investigation of possibilities to
optimize this procedure are in preparation and will be presented
elsewhere \cite{our_work}.  

\begin{figure}
\includegraphics[width=0.95\textwidth]{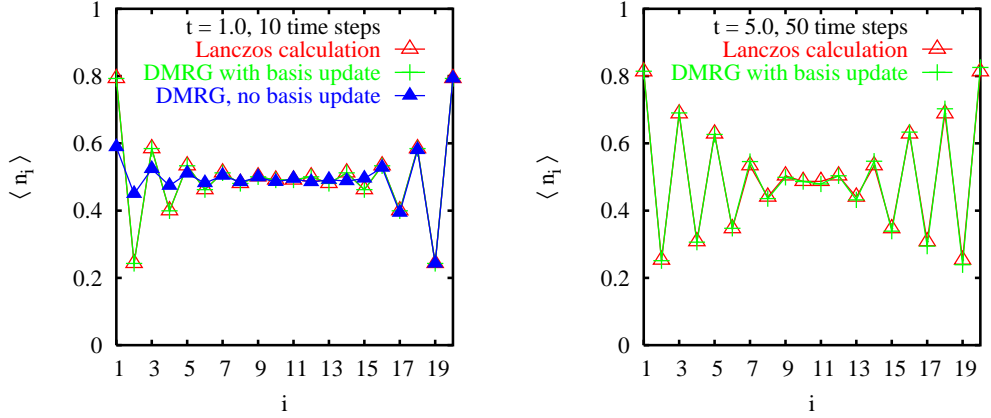}
\caption{Comparison between the time-evolution obtained with the ED
  approach of Sec. \ref{sec::ed}, DMRG without basis adaption, and DMRG
  with basis adaption, as described in Sec.\
  \ref{sec::adaptive_tDMRG} for a system of spinless fermions with 20
  sites, open boundary conditions, and $V(t=0) = 0.5, \, V(t>0) = 2.5,
  \, m_L = 10$.}
\label{fig::compare_exact_dmrg}
\end{figure}

In Fig.\ \ref{fig::compare_exact_dmrg}, we show preliminary results
for a system of spinless fermions with open boundary conditions for
$L=20$.  
As can be seen, the time evolution without basis adaption yields 
a result that is qualitatively wrong even for very small times, while the 
adaptive DMRG retains its accuracy for much longer times.
A comparison of the non-basis-adapted and basis-adapted methods with
our current time-evolution program, which does not
utilize conserved quantum numbers \cite{note_alps},
shows that the error is significant even after only 3 time steps for
the non-adaptive method, while the adaptive method can reach 50 time
steps before the error becomes discernable. 
%When keeping the same number of density matrix basis vectors $m$
%without basis adaption, for the present problem with the present
%program without quantum-numbers, the error is significant already after
%3 time steps, while with the presented basis adaption scheme one can
%reach about 50 time steps before the error has the same magnitude. 

\section{Discussion}
\label{sec::discussion}

In this contribution, we have proposed schemes for the time evolution
of strongly correlated one-dimensional quantum systems based on exact
diagonalization and on the DMRG.
In the ED-based variant, we have applied a Krylov-space approach utilizing
a L\'anczos iterative diagonalization scheme to a system of spinless 
fermions. 
We find that an adiabatic change in the interaction strength induces a
collapse and revival of the metallic state. 
The same system was used to explore possible methods for adaptive
time evolution within the DMRG. 
As discussed in detail, it is crucial to adapt the density-matrix
basis to the changing state at each time step; we have implemented 
such a scheme based
%The adaptive scheme that we have implemented is based 
on a proposal by White \cite{steve_leiden,white_timeevolve_newpaper}. 
%We have compared with the results of the ED scheme.
A comparison of results obtained with the adaptive DMRG scheme with
the ED calculations shows that it can produce accurate results at
fairly long time.
However, a more thorough error analysis and a better understanding of how
to optimize the scheme are necessary. 
%In particular, it is not yet clear how to optimize this scheme. 
Such work is in progress 
%This is under investigation 
and will be presented
elsewhere \cite{our_work}. 

%%%%%%%%%%%%%%%%%%%%%%%%%%%%%%%%%%%%%%%%%%%%%%%%
%% BACKMATTER
%%%%%%%%%%%%%%%%%%%%%%%%%%%%%%%%%%%%%%%%%%%%%%%%

\begin{theacknowledgments}
 We acknowledge useful discussions with S.R.\ White, U.\ Schollw\"ock, and
 C.\ Kollath. We thank the NIC at the Forschungszentrum J\"ulich
 for an allocation of computer time on the IBM-Cluster JUMP. S.R.M.\ 
 acknowledges financial support by the European Graduate College
 ``Electron-Electron Interactions in Solids'' at the Universit\"at
 Marburg. 
\end{theacknowledgments}

%% If the bibliography is produced without BibTeX comment out the

\bibliographystyle{aipproc}   % if natbib is available

\bibliography{vietri_proc}

\begin{thebibliography}{37}
\expandafter\ifx\csname natexlab\endcsname\relax\def\natexlab#1{#1}\fi
\providecommand{\enquote}[1]{``#1''}
\expandafter\ifx\csname url\endcsname\relax
  \def\url#1{\texttt{#1}}\fi
\expandafter\ifx\csname urlprefix\endcsname\relax\def\urlprefix{URL }\fi
\providecommand{\eprint}[2][]{\url{#2}}

\bibitem[Greiner et~al.(2002)]{Bloch}
M.~Greiner, O.~Mandel, T.~W. H\"ansch, and I.~Bloch, \emph{Nature},
  \textbf{419}, 51 (2002).

\bibitem[Rammer and Smith(1986)]{keldysh_review}
J.~Rammer, and H.~Smith, \emph{Rev. Mod. Phys.}, \textbf{58}, 323 -- 359
  (1986).

\bibitem[Berges(unpublished)]{dynamics_QFT}
J.~Berges, \emph{hep-ph/0409233} (unpublished).

\bibitem[Grotendorst et~al.(2002)]{QMC_review1}
J.~Grotendorst, D.~Marx, and A.~Muramatsu, editors, John von Neumann Institute
  for Computing, J\"ulich, 2002.

\bibitem[Nightingale and Umrigar(1999)]{QMC_review2}
M.~P. Nightingale, and C.~J. Umrigar, editors, vol. 525, NATO Science Series,
  1999.

\bibitem[Laflorencie and Poilblanc(2004)]{ED_review}
N.~Laflorencie, and D.~Poilblanc, \emph{Lect. Notes Phys.}, \textbf{645}, 227
  (2004).

\bibitem[White(1992)]{dmrg_prl}
S.~R. White, \emph{Phys. Rev. Letters}, \textbf{69}, 2863 (1992).

\bibitem[White(1993)]{dmrg_prb}
S.~R. White, \emph{Phys. Rev. B}, \textbf{48}, 10345 (1993).

\bibitem[Peschel et~al.(1999)]{dmrgbook}
I.~Peschel, X.~Wang, M.~Kaulke, and K.~Hallberg, editors, Springer Verlag,
  Berlin, 1999.

\bibitem[Schollw\"ock(to appear January 2005)]{uli_rmp}
U.~Schollw\"ock, \emph{cond-mat/0409292; Rev. Mod. Phys.} (to appear January
  2005).

\bibitem[Mak(1992)]{qmc_mak_timeevolve}
C.~H. Mak, \emph{Phys. Rev. Letters}, \textbf{68}, 899--902 (1992).

\bibitem[Egger and Mak(1994)]{qmc_egger_timeevolve}
R.~Egger, and C.~H. Mak, \emph{Phys. Rev. B}, \textbf{50}, 15210--15220 (1994).

\bibitem[Ostilli and Presilla(unpublished)]{QMC_timeevolve}
M.~Ostilli, and C.~Presilla, \emph{cond-mat/0407758} (unpublished).

\bibitem[Luo et~al.(2003)]{luo:049701}
H.~G. Luo, T.~Xiang, and X.~Q. Wang, \emph{Phys. Rev. Letters}, \textbf{91},
  049701 (2003).

\bibitem[Schmitteckert(2004)]{schmitteckert:121302}
P.~Schmitteckert, \emph{Phys. Rev. B}, \textbf{70}, 121302(R) (2004).

\bibitem[Vidal(2004)]{vidal:040502}
G.~Vidal, \emph{Phys. Rev. Letters}, \textbf{93}, 040502 (2004).

\bibitem[White and Feiguin(2004)]{white:076401}
S.~R. White, and A.~E. Feiguin, \emph{Phys. Rev. Letters}, \textbf{93}, 076401
  (2004).

\bibitem[Daley et~al.(2004)]{Uli_timeevolve}
A.~J. Daley, C.~Kollath, U.~Schollw\"{o}ck, and G.~Vidal, \emph{J. Stat. Mech.:
  Theor. Exp.}, p. P04005 (2004).

\bibitem[White(2004)]{steve_leiden}
S.~R. White, \enquote{Methods for Real time Dynamics,} 2004,
  \urlprefix\url{http://komet337.physik.uni-mainz.de/Jeckelmann/DMRG/workshop/%
proceedings/white.pdf}, talk given at the workshop "Recent Progress and
  Prospects in DMRG".

\bibitem[Feiguin and White(unpublished)]{white_timeevolve_newpaper}
A.~E. Feiguin, and S.~R. White (unpublished).

\bibitem[Manmana and Noack(in preparation)]{proceedings_Vietri_DMRG}
S.~R. Manmana, and R.~M. Noack (in preparation).

\bibitem[Wilson(1975)]{wilson_rg}
K.~G. Wilson, \emph{Rev. Mod. Phys.}, \textbf{47}, 773 (1975).

\bibitem[White and Noack(1992)]{white_and_noack}
S.~R. White, and R.~M. Noack, \emph{Phys. Rev. Letters}, \textbf{68}, 3487
  (1992).

\bibitem[Press et~al.(1993)]{numerical_recipes}
W.~H. Press, S.~A. Teukolsky, W.~T. Vetterling, and B.~P. Flannery,
  \emph{Numerical Recipes in C++}, Cambridge University Press, 1993, 2nd edn.

\bibitem[Cazalilla and Marston(2002)]{cazalilla:256403}
M.~A. Cazalilla, and J.~B. Marston, \emph{Phys. Rev. Letters}, \textbf{88},
  256403 (2002).

\bibitem[L\'anczos(1950)]{Lanczos}
K.~L\'anczos, \emph{J. Res. Natl. Bur. Stand}, \textbf{45}, 225 (1950).

\bibitem[Cullum and Willoughby(1985)]{Lanczos_book}
J.~K. Cullum, and R.~A. Willoughby, \emph{Lanczos algorithms for large
  symmetric eigenvalue computations}, vol.~1, Progress in Scientific Computing,
  1985.

\bibitem[Hochbruck and Lubich(1999)]{lubich_timeevolve}
M.~Hochbruck, and C.~Lubich, \emph{BIT}, \textbf{Vol. 39}, pp 620 -- 645
  (1999).

\bibitem[Moler and Loan(2003)]{19dubiousways}
C.~Moler, and C.~V. Loan, \emph{SIAM Review}, \textbf{45}, 3--49 (2003).

\bibitem[Hochbruck and Lubich(1997)]{lubich_errors}
M.~Hochbruck, and C.~Lubich, \emph{SIAM J. Numerical Anal.}, \textbf{Vol. 34},
  pp 1911 -- 1925 (1997).

\bibitem[Gubernatis et~al.(1985)]{spinless_fermions}
J.~Gubernatis, D.~Scalapino, R.~Sugar, and W.~Toussaint, \emph{Phys. Rev. B},
  \textbf{32}, 103 -- 116 (1985).

\bibitem[Voit(1995)]{voit}
J.~Voit, \emph{Rep. Prog. Phys.}, \textbf{58}, 977--1116 (1995).

\bibitem[Manmana et~al.(in preparation)]{our_work}
S.~R. Manmana, A.~Muramatsu, and R.~M. Noack (in preparation).

\bibitem[Suzuki(1976)]{trotter_suzuki}
M.~Suzuki, \emph{Prog. Theor. Phys.}, \textbf{56}, 1454 (1976).

\bibitem[Gobert et~al.(unpublished)]{gobert_timeevolve}
D.~Gobert, C.~Kollath, U.~Schollw\"ock, and G.~Schuetz, \emph{cond-mat/0409692}
  (unpublished).

\bibitem[Kollath et~al.(unpublished)]{corinna_timeevolve}
C.~Kollath, U.~Schollw\"ock, J.~von Delft, and W.~Zwerger,
  \emph{cond-mat/0411403} (unpublished).

\bibitem[not(Our current time-evolution program is based on a simple DMRG
  program developed as part of the ALPS project and utilizes a Lánczos-code
  based on the IETL library, see http://alps.comp-phys.org and
  cond-mat/0410407.)]{note_alps}
 (Our current time-evolution program is based on a simple DMRG program
  developed as part of the ALPS project and utilizes a Lánczos-code based on
  the IETL library, see http://alps.comp-phys.org and cond-mat/0410407.).

\end{thebibliography}

\IfFileExists{\jobname.bbl}{}
 {\typeout{}
  \typeout{******************************************}
  \typeout{** Please run "bibtex \jobname" to obtain}
  \typeout{** the bibliography and then re-run LaTeX}
  \typeout{** twice to fix the references!}
  \typeout{******************************************}
  \typeout{}
 }

\end{document}